\documentclass[sigconf]{acmart}
\usepackage{listings}
\usepackage{xcolor}
\usepackage{enumitem}
\usepackage{booktabs}   
\usepackage{array}      
\usepackage{listings}
\usepackage{multirow}
\usepackage{float}      
\usepackage{afterpage}  
\usepackage{paralist}
\usepackage{graphicx}
\usepackage{tikz}
\usepackage{makecell} 

\renewcommand\footnotetextcopyrightpermission[1]{}
\newcommand*\BC[1]{\tikz[baseline=(char.base)]{
	\node[shape=circle,draw,inner sep=0.15pt] (char) {\textcolor{black}{#1}};}}

\newcommand{\commentout}[1]{}

\author{\rm Jia Hu}
\affiliation{%
\institution{The University of Manchester}
\country{United Kingdom}
}

\author{\rm Youcheng Sun}
\affiliation{
\institution{Mohamed bin Zayed University of Artificial Intelligence}
\country{United Arab Emirates}
}

\author{\rm Pierre Olivier}
\affiliation{
\institution{The University of Manchester}
\country{United Kingdom}
}

\begin{document}

\title{Compartmentalization-Aware Automated Program Repair}

\begin{abstract}
Software compartmentalization breaks down an application into compartments isolated from each other: an attacker taking over a compartment will be confined to it, limiting the damage they can cause to the rest of the application.
Despite the security promises of this approach, recent studies have shown that most existing compartmentalized software is plagued by vulnerabilities at cross-compartment interfaces, allowing an attacker taking over a compartment to escape its confinement and negate the security guarantees expected from compartmentalization.
In that context, securing cross-compartment interfaces is notoriously difficult and engineering-intensive.

In light of recent advances in Automated Program Repair (APR), notably through the use of Large Language Models (LLMs), this paper presents a work in progress investigating the suitability of LLM-based APR at securing cross-compartment interfaces as automatically as possible.
We observe that existing APR approaches and general purpose/code-centric LLMs used as is are unfit for this task, and present the design, implementation, and early results of a new APR framework dedicated to compartment interface safety.
The framework integrates into a feedback loop 1) a specialized fuzzer uncovering cross-compartment interface vulnerabilities; 2) a patch generation component bridging the lack of compartmentalization awareness of existing LLMs with a series of analysis techniques; and 3) a patch validation component assessing the effectiveness of generated vulnerability fixes.
We validate our framework over a sample interface vulnerability, comparing it to a naive use of general-purpose LLMs, and discuss future research avenues.
\end{abstract}

\maketitle
\pagestyle{plain}

\section{Introduction}

Software compartmentalization~\cite{PRIVMAN, PRIVTRANS, lefeuvre2025sok} is a defensive programming practice in which an application is broken down into different software components (compartments) that are isolated from each other, with the goal of limiting the impact of security breaches: an attacker taking over one part of a compartmentalized program will be confined to the containing compartment and unable to access the rest of the application.
Compartmentalization has been demonstrated to protect systems software against memory safety vulnerabilities~\cite{MALICIOUS_DEVICE_DRIVERS}, to provide fault isolation~\cite{LIEDTKE}, to isolate untrusted third-party software and stop supply-chain attacks~\cite{narayan2020retrofitting, BREAKAPP}, to guard against side channels~\cite{SWIVEL}, to hide cryptographic secrets~\cite{SECAGE}, and to isolate unsafe parts of programming languages~\cite{LIBHERMITMPK}.
The effectiveness of compartmentalization at protecting production software has also been proven in the field~\cite{SUCCESS1, SUCCESS2, SUCCESS3}.

The cross-compartment interfaces that result from the compartmentalization of an application represent as many internal trust boundaries that may be used as attack vectors by a malicious actor taking over one compartment and wishing to escape its confinement.
Hence, such cross-compartment interfaces need to be sanitized based on the trust model the programmer aims to enforce through the compartmentalization.
Missing or improper safety checks on the data/control flow at the level of these interfaces lead to Compartment Interface Vulnerabilities (CIVs): flaws enabling cross-compartment attacks that have been shown to negate most of the security guarantees provided by compartmentalization~\cite{lefeuvre2022assessing}.

Securing cross-compartment interfaces and fixing CIVs is both a daunting and difficult task.
It represents a high amount of engineering because CIVs are rampant in today's compartmentalized software~\cite{lefeuvre2022assessing}.
It is also a challenging undertaking, requiring expert knowledge in both the compartmentalized application and in security-aware software engineering practices.
Addressing CIVs requires a deep understanding of the application's intended behavior, familiarity with its software architecture and design patterns, and the ability to identify the root cause of the failure.
This process also requires expertise in defensive programming: developers must manually locate the violated compartment trust boundaries and implement the appropriate data and control flow safety checks.
Because of these difficulties, existing works often scope out or neglect interface safety~\cite{lefeuvre2022assessing}.
All in all, CIVs have been identified as one of the main roadblocks towards a wider adoption of compartmentalization as a standard software engineering practice~\cite{lefeuvre2025sok}, effectively preventing a broader audience from reaping the security benefits that can be gained from this practice.

Automatic Program Repair~\cite{weimer2009automatically, APR, zhang2023critical} (APR) is a software engineering practice aiming at automatically addressing software bugs without the need for human effort or expertise.
Recent advances in artificial intelligence have led to a surge of interest in applying these techniques to APR.
In particular, the use of large language models (LLMs) for APR~\cite{bouzenia2024repairagent} has emerged as a promising direction.
In that context we ask the following question: \emph{can LLM-based APR approaches help reduce the engineering effort of securing cross-compartment interfaces in compartmentalized software?}

This paper presents a work in progress investigating that research question.
We observe that existing approaches at LLM-based APR cannot be applied as-is, because of several limitations that represent as many challenges for our study and can be summed up as a lack of awareness of compartmentalization.
Existing generalist or code-focused LLMs are mostly trained on monolithic, non-compartmentalized software, and the lack of a large set of existing compartmentalized software makes it difficult to train domain-specific models.
In this context, current APR approaches and the relevant LLMs are unaware of key aspects of compartmentalized software including trust boundaries and relationships between the compartments forming an application, as well as the nature of the control and data flow involved in the exploitation of a CIV.
As a result, when faced with a CIV to fix, existing APR approaches struggle to identify what a fully secure fix should be, and where it should be located in the target application's code.

To address this gap and understand to what extent LLM-based APR can automate securing interfaces and addressing CIVs, this paper presents a work in progress in which we design and implement a compartmentalization-aware repair framework for CIVs that makes trust assumptions explicit and uses them to guide repair decisions.
The framework takes as input a particular interface within a compartmentalized application, and its goal is to automate the securing of this interface as much as possible.
To that aim the framework make use of general purpose LLMs to drive the repair process.
When faced with a CIV to fix, the compartmentalization knowledge gap these models suffer from is filled by feeding them extensive information about the vulnerability and the considered compartmentalized application.
That information includes in particular the output of two CIV analysis techniques we propose.
These techniques classify CIVs based on the data type(s) involved in the payload and model the payload's flow on the execution call stack to drive the LLM to produce the best patch, and the best location in the target application's code.
The framework integrates an in-memory fuzzer that searches for and produces CIVs at the level of the considered interface for the APR component to fix.
The fuzzer also act as part of a feedback loop with the APR component, being leveraged to validate the patches proposed by the APR and refine them to avoid fixing CIVs only partially.

Our future evaluation plans include comparing the effectiveness of our approach to a baseline, the naive use of LLMs to attempt to fix CIVs, and to competitors that are existing compartmentalization-unaware APR frameworks, using an existing database of CIVs~\cite{lefeuvre2022assessing}.
Beyond helping to make compartmentalized software more secure, this study will help qualify and quantify how good are LLM-based APR approaches are tackling CIVs, what are the types of CIVs that can be fixed this way and what are the types still requiring human intervention.
For the latter, the study will also investigate how can the APR framework guide developers in the fixing of CIVs that cannot be fully addressed automatically.

This paper makes the following contributions: 1) we design and implement \emph{an LLM-based APR framework dedicated to securing cross-compartment interfaces by addressing CIVs}. The framework uses an LLM for patch generation, made compartmentalization-aware through the prompt, integrated with a fuzzer into a feedback loop; and 2) we present \emph{two analysis techniques that classify CIVs} based on the type of the payload involved and the different functions composing the execution call stack where the payload flows. These classifications are used to guide the LLM to generate the most suitable patches at the best locations in the target application.

\section{Background}

\subsection{Software Compartmentalization and CIVs}

\paragraph{Software Compartmentalization.}

Software compartmentalization applies the principle of least privilege~\cite{LEAST_PRIVILEGE} at the granularity of program components~\cite{lefeuvre2025sok}. A program is decomposed into compartments that execute with limited privileges and are isolated from one another so that compromise of one component is less likely to endanger the rest of the application. One common instantiation places each compartment (e.g., individual libraries) in a separate process. Isolation is then enforced by hardware or OS mechanisms such as page tables, which restrict unauthorized control flow and data access across compartments. The objective is to preserve the confidentiality, integrity, and/or availability of non-compromised compartments even when another compartment is subverted. At the same time, compartments within the same application must still communicate, and therefore interact through explicit interfaces, such as inter-process communication channels, that ideally validate and sanitize cross-compartment inputs.

Compartmentalization has been used to mitigate a broad of threats and faults~\cite{MALICIOUS_DEVICE_DRIVERS, LIEDTKE, narayan2020retrofitting, BREAKAPP, SWIVEL, SECAGE, LIBHERMITMPK}, and has demonstrated its effectiveness at accomplishing this task in production~\cite{SUCCESS1, SUCCESS2, SUCCESS3}.
Examples of production-ready compartmentalized software include web browsers~\cite{SITE_ISOLATION, narayan2020retrofitting}, web servers~\cite{APACHE}, operating system~\cite{LIEDTKE}, or various system utilities~\cite{goldberg1996secure}.

\paragraph{Trust Models.}

\begin{figure}
    \center
    \includegraphics[width=0.45\textwidth]{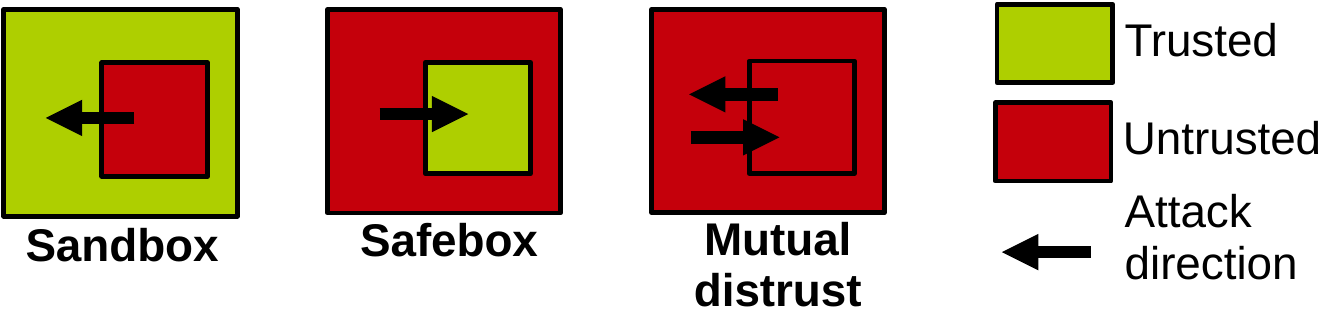}
    \caption{Trust models enforced by software compartmentalization, with the direction of attacks exploiting CIVs by corrupting the control or data flowing between untrusted and trusted compartments.}
    \label{fig:trust-models}
\end{figure}

Compartmentalization can enforce 3 trust models~\cite{lefeuvre2025sok}, presented on Figure~\ref{fig:trust-models}.
With the \emph{sandbox}~\cite{goldberg1996secure} model, a compartment (e.g., third-party code) is untrusted and as such isolated from the rest of the application.
With the \emph{safebox} model, a compartment (e.g., code manipulating secrets) is critical to the security of the application and needs to be isolated from the rest of the application.
Finally, \emph{mutual distrust} is a generalization of the two previous trust models and represents compartments distrusting each other, e.g., as is the case between the OS and a trusted application within a trusted execution environment.
The examples illustrated on Figure~\ref{fig:trust-models} generalize to applications with more than 2 compartments with possible combinations of trust models.

Compartmentalization assumes that untrusted compartments may become malicious.
Should that happen, an attacker taking over a compartment will aim to escape the confinement enforced by compartmentalization to access the rest of the application, as represented by the arrow labeled \emph{attack direction} on Figure~\ref{fig:trust-models}.
Direct access is generally not possible due to the use of an isolation mechanism, however the attacker can still try to misuse legitimate communication channels between the compromised compartment and the rest of the application, with the hope of triggering a vulnerability allowing them to escape the compartmentalization.

\paragraph{Compartment Interface Vulnerabilities.}

Compartment interface vulnerabilities (CIVs)~\cite{lefeuvre2022assessing} arise at cross-compartment interfaces due to the lack of sanitization on the data and control flow between untrusted and trusted compartments.
Using a CIV, the isolation enforced by compartmentalization can be bypassed by a malicious compartment misusing the legitimate communication interfaces it has with other compartments.
This can be achieved through the injection of malformed data, for example a malicious compartment may call a function exposed by its victim and passing it malformed parameters or pointers to malformed data in shared memory if there is any.
A victim compartment may also call a function exposed by a malicious one, which responds with corrupted data in the return value or shared memory.
The legitimate cross-compartment control flow intended by the programmer in a victim compartment can also be altered by a malicious one by calling functions exposed by the victim in the wrong order, installing malicious callbacks, or corrupting shared synchronization primitives.

CIVs have been classified~\cite{lefeuvre2022assessing} into 1) cross-compartment data leakages such as exposure of secrets or addresses; 2) cross-compar\-tment data corruption such as corrupted pointers, indexing information, or data structures; and 3) temporal violations such as improper order of API calls, corrupted synchronization primitives, or time-of-check-to-time-of-use attacks on cross-compartment shared memory.
These allow attackers to break the cross-compartment confidentiality, integrity, and/or availability that compartmentalization aims to enforce, which negates its security benefits.

CIVs exist because of a lack of safety checks on both the data flowing into a trusted compartment from outside, and on the API it exposes.
CIVs have been shown to be prevalent in compartmentalized software.
When compartmentalization is retrofitted into an existing monolithic application, the interfaces that emerge between the newly created compartments represent as many trust boundaries that need to be secured, something that represents a very difficult task given that the application in question was never designed with internal trust boundaries in mind.
If fact, the vast majority of existing compartmentalization research works at best scope out or even ignore the issue of sanitizing interfaces~\cite{lefeuvre2025sok}.
Even when software is designed from scratch with compartmentalization in mind, fully securing cross-compartment interfaces is still extremely hard.
The recent explosion of popularity for fuzzing is a testament to the difficulty of properly securing interface between distrusting software components.

\begin{lstlisting}[float,
  caption={CIV in FFmpeg with libavcodec sandboxed. In this scenario libavcodec runs within an untrusted compartment and returns a \texttt{NULL} pointer \texttt{cfg} to FFmpeg's (trusted) compartment. That pointer is then dereferenced without being checked, giving the libavcodec's compartment the ability to crash FFmpeg's compartment, compromising its availability.},
  label={lst:ffmpeg}]
static void print_all_libs_info(int flags, int level) {
    // simplified version of ffmpeg code

    const char *cfg = libavcodec_configuration(); // returns NULL
    if (strcmp(FFMPEG_CONFIGURATION, cfg)) {      // dereferences NULL
        // ...
    }

    // ... more code to print other libraries' info
}
\end{lstlisting}

Listing \ref{lst:ffmpeg} presents a simple example of CIV in FFmpeg with libavcodec sandboxed.
We have two compartments: the untrusted libavcodec sandbox, and the rest of the FFmpeg program's code, which is trusted.
The code snippet is an excerpt of the FFmpeg compartment's code invoking a function exposed by the libavcodec compartment, \texttt{libavcodec\_configuration}.
That function returns a pointer \texttt{cfg} which is then passed to \texttt{strcmp} without any form of check and dereferenced.
In a scenario where the untrusted libavcodec has been taken over by an attacker, which we assume gained arbitrary code execution capabilities within the context of the corresponding compartment, \texttt{libavcodec\_configuration} may return a \texttt{NULL} pointer allowing the subverted libavcodec compartment to crash the FFmpeg compartment and bypass compartmentalization to compromise availability.

Addressing this CIV would require to implement checks on the trusted side (FFmpeg), ensuring that the pointer \texttt{cfg} reference valid (mapped) memory, and that the string referenced is well terminated to avoid an overflow in \texttt{strcmp}.
Fallback code in case that check fail may also need to be implemented.

\section{Approach}

\subsection{Threat Model}

We assume a compartmentalized application written in a memory-unsafe language and made of two or more compartments, as well as an attacker armed with the ability to subvert an untrusted compartment through a traditional vulnerability.
We assume a strong attacker able to gain arbitrary code execution capability within the context of the subverted compartment.
The attacker's goal is to bypass the compartmentalization and compromise the confidentiality, integrity, and/or availability of the compartments outside the one subverted.
We also assume that the interface between the subverted compartment and the rest of the application suffers from CIVs, which the attacker aim to exploit to reach their goal.
Another assumption we make is that the attacker only attempts to trigger CIVs by misusing the data flowing between compartments, which represents the vast majority of CIVs~\cite{lefeuvre2022assessing}, and we scope control flow-based cross-compartment attacks as future works.
We also only consider CIVs involving one malicious and one victim compartment, a prevalent scenario vs. more complex ones such as having several compartments under the control of the attacker.

In this context we aim to create an APR framework that can address as many CIVs as possible, as automatically as possible, to guide the developer in implementing the relevant fixes to sanitize interfaces between untrusted compartments and trusted ones.

\subsection{Objectives}
We aim to design and implement an iterative LLM-based APR framework to secure interfaces in compartmentalized applications and address CIVs.
Our goals are both to improve the security of compartmentalized software, and assess the effectiveness LLM-based APR at such a task.
To that aim we identify the following objectives and associated challenges.

\paragraph{Compartmentalization-Aware Reasoning.}
Our framework secures an interface one CIV at a time.
When working on fixing a particular CIV for a given interface, the patches generated by the framework need to align with the security objectives of the compartmentalized application in question.
To that aim the LLM must be made aware of \emph{important characteristics of the compartmentalized application and the interface considered}: what the compartments interacting through the interface, what and where are the corresponding trust boundaries between them, what are the trust relationships between the compartments involved, what are the security goals of the specific compartmentalization policy (confidentiality, integrity, and/or availability), etc.
The LLM must also be made aware of \emph{important characteristics of the CIV considered:} how does the victim compartment reacts when it is triggered (e.g. fault in read or write mode), where is the fault located and on what call stack, where is the payload injected from, what are the data structures involved, etc.
LLMs are seldom trained on compartmentalized software or security interfaces, and many of the aforementioned characteristics are specific to the applications, interfaces, compartmentalization policies, and CIVs considered.
These characteristics must then be extracted from descriptions of the use case (e.g., fuzzing or crash logs) considered through analysis and fed to the LLM via the prompt, in order to drive the APR to generate the best fixes at the best possible location in the code.

One of our goals is also to identify what are the types of CIVs that can be efficiently fixed by LLM-based APR, and what are the types with which our approach struggles.
To that aim, we need a classification of CIVs, that should also prove useful in driving the LLM to produce patches.

\paragraph{Access to Large Amounts of CIVs.}

One of our goals is to classify what CIVs can be addressed by LLM-based APR and what CIVs cannot, hence we need a large set of CIVs in order to be able to take generalizable conclusions.
To that aim a central component of our framework is a CIV fuzzer that can uncover interface vulnerabilities.
Here we use the ConfFuzz~\cite{lefeuvre2022assessing} fuzzer to accomplish that task.
ConfFuzz can not only extract CIVs by fuzzing compartmentalized software, but can also fuzz monolithic software at potential (e.g., libraries) compartment boundaries and extract the CIVs these applications would suffer from if compartmentalized without sanitizing interfaces.
As a result ConfFuzz lets us access a vast amount of CIVs, including the 629 CIVs from the original ConfFuzz study~\cite{lefeuvre2022assessing}.

\paragraph{Iterative Patch Generation and Validation.}

As it is often the case with LLM-based APR, we anticipate that many patches generated will not fix the CIV in question.
We also anticipate that certain patches will solve a CIV only partially, for example a \texttt{NULL} pointer check may avoid certain crashes but will still fault when dereferenced if its value corresponds to any non-\texttt{NULL} but unmapped/inaccessible area of memory.
The patch generation component of our framework must then be coupled with a validation method that assess the effectiveness of the patch at mitigating the considered CIV, as well as variations of that particular CIV.
That validation component must operate in a feedback loop with the patch generation component, validating and increasing the quality of patches iteratively.
If it is determined that a CIV cannot be fixed or fully fixed, a report should be produced describing the issue and guiding a human programmer as much as possible to address the CIV manually.

\subsection{Overview}

\begin{figure}
  \center
  \includegraphics[width=\linewidth]{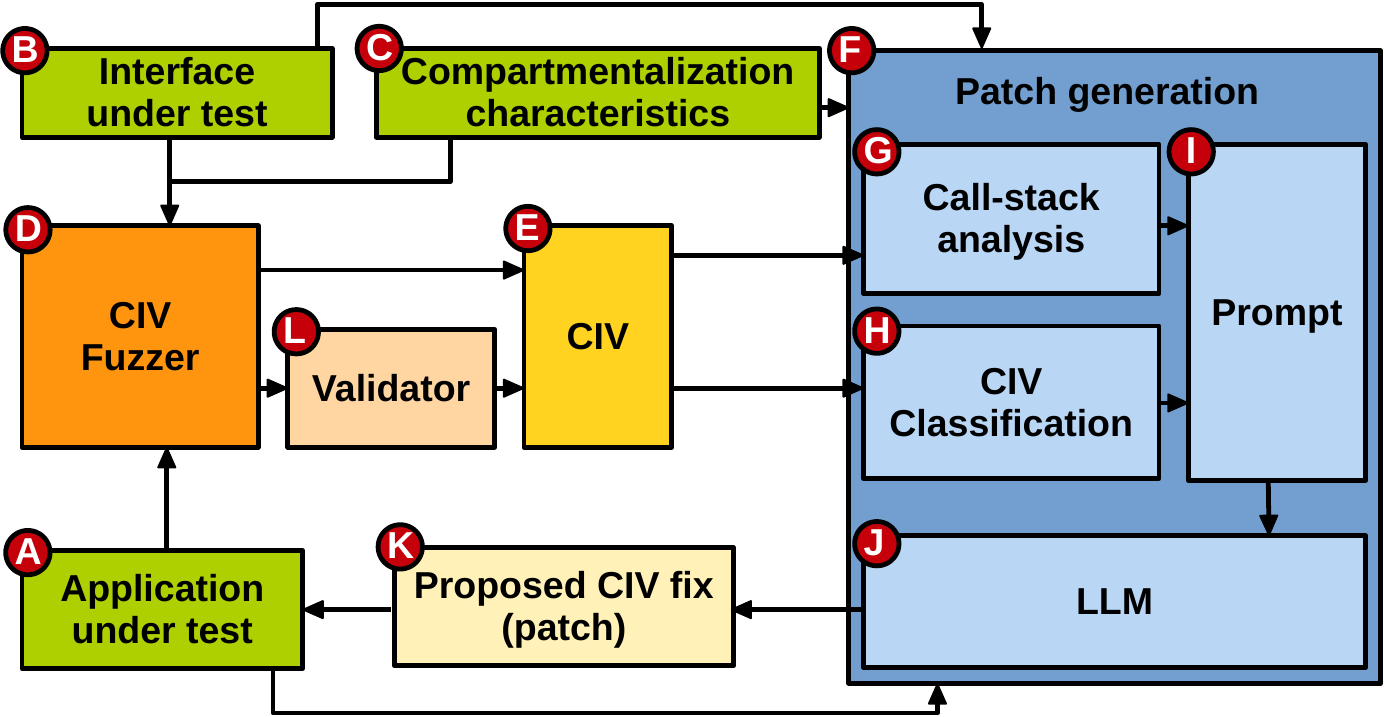}
  \caption{Overview of our compartmentalization-aware APR framework. It takes as inputs (green boxes) an application's source code, a description of its desired compartmentalization policy and of a particular cross-compartment interface to secure. A CIV Fuzzer uncovers CIVs on that interface, and these CIVs are fed to the patch generation component. That component analyzes the CIV and other inputs to construct a prompt for a LLM to generate a candidate patch addressing for the CIV. That patch is then validated by invoking the fuzzer again to check if the CIV is properly fixed. The process iterates until the CIV is fully addressed, or the framework determines it cannot do so.}
  \label{fig:overview}
\end{figure}

Figure~\ref{fig:overview} presents an overview of our framework, that we develop on in the next paragraphs.

\paragraph{Inputs}

The framework takes as inputs the source code of the compartmentalized application under test~\BC{A}; a description of the interface to secure~\BC{B} which is simply a list of functions that make up the boundary between the considered compartments; and a description of the compartmentalization characteristics~\BC{C} which consists in the compartmentalization policy, i.e. what parts of the application's code go into what compartments, and the trust model: which compartment is considered untrusted/malicious.
As discussed earlier we limit our study to cases with a single malicious compartment, and consider the rest of the application as trusted, i.e., as potential targets for a CIV-based attack.

\paragraph{Fuzzing to Uncover CIVs}

The framework's inputs are first processed by a CIV fuzzer~\BC{D} which builds the application after having instrumented the untrusted side of the interface considered in order to mutate the data flowing through it.
The goal is to inject malformed payloads onto the trusted side through that interface and detect CIVs.
The application is also instrumented with sanitizers (e.g., ASan) to maximize the chances of detecting such vulnerabilities.
We use ConfFuzz as the fuzzing element of our framework and refer the reader to the relevant paper~\cite{lefeuvre2022assessing} for details about its inner workings.
When a CIV~\BC{E} is uncovered by the fuzzer, it outputs a sanitizer report which is passed along with the other framework inputs (application source code, interface to secure, compartmentalization characteristics) to the patch generation component~\BC{F}.

\paragraph{LLM-Based APR}

The patch generation is the core element of our APR framework: armed with a CIV, its goal is to prompt a LLM to generate a patch addressing the CIV.
To that aim, as discussed previously, the prompt~\BC{I} must be constructed to inform the LLM~\BC{J} as much as possible about the compartmentalized scenario considered.
Key information that goes into the prompt include 1) \emph{details about the compartmentalized application}: its code, compartmentalization policy, and the trust model enforced; and 2) \emph{details about the CIV}: where and how it occurs (sanitizer report) and how it was triggered (fuzzer mutation and instrumentation log).
Regarding the latter, we characterize the nature of the CIV by analyzing the sanitizer report and the application source and classifying~\BC{H} the vulnerability into a series of categories that will be described in details in Section~\ref{sec:civ-classification}.
We also perform an analysis of the execution call stack A) at the time the payload was injected from the malicious compartment and B) at the time the fault it led to was triggered on the trusted side~\BC{G}.
This analysis attempts to recover the data flow path of the payload and help the LLM find the best location of the CIV's fix, as we describe in Section~\ref{sec:stack-classification}.
The output of both analyses is used to further enhance the prompt.
The LLM outputs a patch~\BC{K} which is a candidate fix for the CIV.

\paragraph{Patch Validation}
To validate the patch~\BC{L} we first launch the fuzzer to reproduce the payload that led to the discovery of the CIV.
To that aim we have updated ConfFuzz to offer a new CIV reproduction feature: during the initial fuzzing process, ConfFuzz uses CRIU~\cite{CRIU} to checkpoint the state of the application before each payload injection.
The snapshot taken before the injection of the payload that triggered the CIV discovery can then be used for reproducing the issue.
If the CIV is still present, the patch is dropped and a new round of patch generation starts, in which the prompt is updated to include information about the failed patch.

Early results showed that the patch generation component will often output a patch addressing a CIV \emph{only partially}.
For example, a fault involving an array being addressed with an index (payload) that is negative can be avoided by checking that the index is superior or equal to 0 before indexing the array.
However, this fix is not comprehensive because it does not handle other occurrences of that CIV in which the payload is a positive index which value is larger than the size of the array in question.
To determine if the fix brought by a patch is partial, our validator component mutates the payload in various ways, attempting to trigger other occurrences of the CIV considered.
If the fix is deemed to be only partial, the framework initiates another round of patch generation with an updated prompt to make the LLM aware of the partial nature of the previous fix.

The validation and the patch generation components work together in a feedback loop.
After a configurable number of iterations, if the CIV is determined to be not fixed or not fully fixed, the framework stops and outputs a report aiming to guide a human developer to address the problem manually.

\subsection{CIV and Call Stack Analysis}

To guide our APR in recognizing the CIV's background threat model, so it can identify reasonable patching sites and generate high-quality refinements, the policy should be applied so that the APR can:
\begin{compactenum}
\item Find the appropriate granularity when scanning the data flow across each compartment.
\item Reason about the role of the functions within the crash stack.
\item Determine the priority for patching by analyzing the interactions between each function.
\end{compactenum}

\subsubsection{CIV Classification}
\label{sec:civ-classification}



\paragraph{Type-Centric Classification}

\begin{table*}[ht]
\caption{Classification of CIVs by data type.}
\footnotesize
\centering
\label{tab:CIV_classification}
\begin{tabular}{l l l l}
\toprule
\textbf{Type} & \textbf{Boundary Focus} & \textbf{Security Impact} & \textbf{Patching Unit} \\
\midrule
Pointer & Legality & $\text{Availability} \gg \text{Integrity}$ & pointer instance \\
Scalar & Domain & $\text{Integrity} \gg \text{Availability}$ & scalar token \\
Structured Payload & Semantic fields & $\text{Confidentiality} \approx \text{Integrity} \gg \text{Availability}$ & field \\
Opaque Handle & Handle validity \& semantics & $\text{Confidentiality} \gg \text{Integrity}$ & handle instance \\
\bottomrule
\end{tabular}
\end{table*}


In order to reason more precisely about the questions of patchability raised earlier, we adopt a type-centric taxonomy as summarized in Table \ref{tab:CIV_classification}. 
We categorize CIVs based on the data types which may ultimately define the boundary semantics at interface-level interactions. We focus on the following types: pointers, scalar values, structured payloads, and opaque handles.

For each data type, we assume the existence of a hypothetical oracle function that performs the strongest possible legality or validity checks for that type, without regard to performance overhead. This assumption enables us to evaluate whether a given CIV is theoretically detectable or fixable under ideal conditions, and therefore to quantify the upper bounds of patchability for each category.

\paragraph{Pointer Types}

For pointer types, we do not consider their semantic meaning; instead, we only evaluate their legality. In other words, we are concerned with the reference itself rather than the object being referenced. We assume the existence of an oracle function, \texttt{is\_pointer\_mapped}, capable of detecting whether a given pointer refers to mapped memory. Such a function can reliably validate the dereferenceability of any address. By integrating this oracle into the generated patches, we can evaluate the extent to which LLM-based patching can resolve this class of CIVs. When pointer-type CIV is exploited, mostly it will influence availability rather than integrity regarding the security impact when considering at the boundary level, and its legality is highly detectable without considering its semantic meaning. Meanwhile, APR should focus on the granularity of specific "pointer instance" at the interface level during patching process.

\paragraph{Scalar Values}

For scalar value types, we assume the existence of an oracle function \texttt{is\_in\_domain} that verifies whether an enumeration value or flag remains within its valid domain. However, a mutated value that is syntactically valid but semantically incorrect cannot be detected, as no intent related constraints are imposed on such data types. When scalar kind of CIV is being exploited at boundary level, the system may still be "up" (available), but it is just doing the wrong things, so the integrity impact is higher than availability when considering security impact. Also, it is detectable if a scalar type variable is outside its legal domain, and during the patching process the APR should focus on the granularity of particular "scalar token" at the interface level.

\paragraph{Structured Payloads}

For structured payload types, although the data may also be passed by pointer, the relevant boundary is not the pointer value itself, but the semantic boundaries of the object it references. These boundaries correspond to multiple semantic interfaces embedded within the structure. In this case, the oracle function may be regarded as a hybrid of the aforementioned checks, depending on the actual data types of the structure’s subfields. Since a data structure is the minimal semantic unit within an application, when this kind of CIV is being exploited, both the confidentiality and integrity are highly impacted rather than availability. With mechanisms such as CHERI or PAC, the memory safety and pointer integrity can be guaranteed, but the detectability is medium since its semantic meaning is transparent across all the syntactic/format-based checking processes. The APR should use "field," which corresponds to the relevant data structure, as the granularity at the interface level during the patching process.

\paragraph{Opaque Handles}

For opaque handle types, we assume an oracle function \texttt{verify\_format} that validates format constraints for a given handle. Nevertheless, the detectability is medium since neither the intent nor the semantic correctness of such objects can be verified solely through format checking. The whole object referred to by the handle will be compromised when this kind of CIV is being exploited; the security impact will mainly focus on confidentiality rather than integrity, and the granularity of the patching process should be implemented via the "handle instance" at cross-compartment interactions.

We categorize CIVs with respect to boundary focus and patching unit under confidentiality, integrity, and availability (CIA) considerations.
It should be noted that CIVs are not always "fixable" in a semantic sense, and often we cannot reconstruct the original correct value.
We can only detect violations and contain their consequences to preserve CIA invariants. Subsequently, we evaluate the effectiveness of the LLM in addressing specific types of CIVs in Section~\ref{sec:case-study}.

\subsubsection{\textbf{Crash Stack Function Classification}}
\label{sec:stack-classification}



\paragraph{Function Classification}

To better differentiate the distinct role of each function within the crash stack, and to better evaluate the efficiency of different patching sites, we provide the following definitions to categorize them according to their semantic contributions with respect to the crash.
For the nondeterministic return time of the sandbox function, we may have a set of candidate functions that have the crash-relevant variable in scope, and are executed after the sandbox function returns, we call them boundary candidates.
After the set of boundary candidates is determined, we can rank each function according to its semantic interaction level with the crash variable to further pick the final boundary function:

\begin{list}{\labelitemi}{\leftmargin=1em}
\item  \textbf{Boundary function:} the earliest trusted function that has the crash-relevant variable in scope and semantically interacts with it. Boundary function may serve as patching target.

\item  \textbf{Crash site:} crash site is the function where the crash actually happens; sometimes it is a library interface without source code.

\item  \textbf{Crash variable:} the crash variable is the concrete instance of data that the attack vector eventually reaches at the crash site.
\end{list}

\paragraph{Function Ranking Process}

The next step is to find the ideal patching site given all the above identified functions, starting earlier from the boundary function, or ending up with the crash site function
There is no deterministic way to make it certain, the mutation may break many areas within the data structure and eventually manifest the crash at the crash site function, other functions within the scope may just be lucky not to touch the broken part, so this time we need to focus on the crash variable.
To implement this, we add a ranking process to find out which function is the earliest one to interact with the crash variable.
We classify the functions composing the crash call stack into the following categories:

\begin{list}{\labelitemi}{\leftmargin=1em}
\item \textbf{CONSUME:} the first function within the boundary candidates that consumes the crash variable.
\item \textbf{FORWARD:} functions within the boundary candidates that pass through the crash variable without interpreting it.
\item \textbf{PRESENCE:} functions within the boundary candidates that have the crash variable in scope but do not interact with it.
\item \textbf{COMMIT:} the function installs the crash variable into trusted structures. It corresponds to untrusted code and will not be one of the candidates.
\end{list}

\label{sec:priority}
Using this model several patching location strategies can be expressed.
Currently, we instruct the LLM to prioritize the patch target location as follows: \textbf{CONSUME} (most prioritized), \textbf{FORWARD}, \textbf{PRESENCE} (less prioritized).
However, when the \textbf{CONSUME} is a library function without available source code, we instead trace upward to the last function that forwards the crash variable (and has source code). We treat this \textbf{FORWARD} function as the patching site in place of the library \textbf{CONSUME} function. This aims to prevent unnecessary checks by patching as late as possible, however other strategies can be applied e.g., patching as early as possible on the trusted code side.


\section{Case Study}
\label{sec:case-study}

This section illustrates how our framework repairs a representative CIV using the approach shown in Figure~\ref{fig:overview}. Guided by the policy-driven prompt, the framework targets the trusted domain, applies our CIV classification, and selects an appropriate level of granularity to identify a CIV-specific patch site. We also compare our approach against a naive GPT-style repair baseline and show improved efficiency in repairing the same CIV.

\subsection{A CIV Under a Sandbox Scenario}

\begin{lstlisting}[float,
  caption={CIV in Apache with \texttt{mod\_markdown} sandboxed. A corrupt pointer is being passed as \texttt{apr\_table\_t *table}, and the crash happens when it is dereferenced by \texttt{apr\_table\_get}.},
  label={lst:apache}]
static int log_table_entry(const apr_table_t *table, const char *name, char *buf, int buflen) {
    // ...
    if ((value = apr_table_get(table, name)) != NULL) //crash site
    // ...
}
\end{lstlisting}

As shown in Listing~\ref{lst:apache}, the crash occurs in a sandbox scenario. The trusted function \texttt{log\_table\_entry} forwards a corrupted \texttt{apr\_table\_t} pointer to \texttt{apr\_table\_get}, which dereferences it and triggers the crash.

The host application is Apache HTTPD. We sandbox the shared library \texttt{mod\_markdown} within it, and by instrumenting this library, we inject malformed data through the sandbox function as shown in Listing~\ref{lst:sandbox-function}.
\begin{lstlisting}[float,
  caption={Sandbox function \texttt{markdown\_output}},
  label={lst:sandbox-function}]
int markdown_output(MMIOT *doc, request_rec *r, markdown_conf *conf)
\end{lstlisting}
We mutated a subfield of the second parameter \texttt{request\allowbreak\_rec *r} to see if this compromised sandbox library can crash the context-side application (Apache HTTPD), the crash will not appear until the corrupted subfield within \texttt{request\_rec} is consumed by context-side (Apache HTTPD) functions. This mutated subfield of data structure (\texttt{request\_rec}) is later dereferenced by the function \texttt{log\_table\_entry}, and finally leads to the crash shown in Listing~\ref{lst:apache-crash-stack}.
\begin{lstlisting}[float,
  caption={Apache HTTPD crash stack.},
  label={lst:apache-crash-stack}]  ...
    #0 0x7fffe1cc6375 in apr_table_get (/usr/lib/x86_64-linux-gnu/libapr-1.so.0+0x1a375)
    #1 0x5555556b1776 in log_table_entry
    #2 0x5555556b18c5 in log_header
    #3 0x5555556b2c9e in do_errorlog_default
    #4 0x5555556b4ea1 in log_error_core
    #5 0x5555556b5706 in ap_log_rerror_
    #6 0x5555556a1d7b in ap_invoke_handler
    #7 0x5555556f5835 in ap_process_async_request
  ...
\end{lstlisting}
The ideal fix for this CIV is to check the integrity of the data structure \texttt{apr\_table\_t} before forwarding it to \texttt{apr\_table\_get}.

\subsection{Compartmentalization-Aware Framework}
Given the crash stack, the framework reasons about the role of each function and applies our ranking policy to select a boundary-relevant patch site before generating the corresponding repair.
In this example, the vulnerability is best characterized as a structured-payload CIV: although the tainted value is passed as a pointer, it originates from a corrupted subfield within a larger interface structure and is later consumed in trusted code.
Applying the boundary-candidate rules to the crash stack yields a small set of candidate functions, corresponding to frames \#1--\#6 in Listing~\ref{lst:apache-crash-stack}.

\paragraph{Boundary Function Selection.}
The patch site is selected by applying our ranking policy to the candidate set and assigning each function a role in the propagation and consumption of tainted data.
As discussed earlier, our patching unit is the fields within data structure. However, since the crash variable is a data structure that is nested inside a larger data structure, the question becomes: at which structural layer should the patch be introduced? The answer lies at the CIV interface mutated by the sandbox, namely the \texttt{request\_rec} structure, we need to find out which function consumes the tainted subfield of this data structure at interface level.

The tainted subfield within \texttt{request\_rec} is \texttt{apr\allowbreak\_table\_t *}, since the crash occurs when \texttt{apr\_table\_get} dereferences this pointer. Accordingly, \texttt{apr\_table\_t *} is used as the repair \emph{unit}, and the boundary-candidate functions are traversed to determine where this unit is first interpreted in trusted code. In this stack, \texttt{log\_table\_entry} is the only candidate that both directly handles the unit at the interface level and is available for modification (i.e., has source code).

\commentout{
\begin{list}{\labelitemi}{\leftmargin=1em}
\item \textbf{UNIT:} the data structure \texttt{apr\_table\_t}.
\item \textbf{CONSUME:} \texttt{apr\_table\_get}, the first function within the boundary candidates interacting with the UNIT, should be the ideal patching target, but since it is a library function without source code, we need to trace upward to a forwarder that has source code.
\item \textbf{FORWARD:} \texttt{log\_table\_entry}, the last forwarder has source code, which is the patching site.
\item \textbf{PRESENCE:} other functions in the boundary candidate set, implicitly carry the UNIT in their data flow but do not touch or interpret it.
\item \textbf{COMMIT:} \texttt{markdown\_output}, the sandbox function that installs the UNIT into trusted memory and thus commits it.
\end{list}
}
\begin{list}{\labelitemi}{\leftmargin=1em}
\item \textbf{UNIT:} the interface-relevant value \texttt{apr\_table\_t *}.
\item \textbf{CONSUME:} \texttt{apr\_table\_get} is the first consumer of the unit, but it is a library function without modifiable source code.
\item \textbf{FORWARD:} \texttt{log\_table\_entry} is the last forwarder with modifiable source code and is therefore selected as the patch site.
\item \textbf{PRESENCE:} other candidates propagate the unit but do not interpret or validate it.
\item \textbf{COMMIT:} \texttt{markdown\_output} is the sandbox entry point that introduces the tainted unit into trusted execution.
\end{list}


\begin{lstlisting}[float,
  caption={HTTPD fix proposed by our framework.},
  label={lst:apache-fix}]
static int log_table_entry(const apr_table_t *table, const char *name, char *buf, int buflen) {
  // ...
  // Check table's validity before dereferencing it
  int table_mapped = is_pointer_mapped((const void *)table, sizeof(*table));
  if (!table_mapped) 
    return -1;
  // ...
  if ((value = apr_table_get(table, name)) != NULL)
  // ...
}

\end{lstlisting}

Now let us examine the final generated patched function. As discussed earlier, we assume the existence of an oracle function \texttt{is\_pointer\_mapped} that checks the legality of a pointer. In this case, the mutation changed the pointer stored in \texttt{apr\_table\_t *table} from a valid value to an unmapped value, which subsequently triggered the crash. To prevent this, we use \texttt{is\_pointer\allowbreak\_mapped} to implement a sanity check prior to dereferencing this field. The patched function becomes as presented in Listing~\ref{lst:apache-fix}.
As shown, we precisely identified the function that holds the authority to operate on the boundary unit, It is the earliest interface-level consumer that can prevent the propagation of the corrupted pointer.

The resulting patch is boundary-interface-specific: the exact remediation strategy depends on how the boundary-level interface interacts with the boundary unit and how the crash is triggered within it. The flexibility lies in the granularity of the boundary unit. Since this patching strategy still needs to be validated against numerous CIVs, we can update it later by adjusting the granularity of the boundary unit.

\subsection{Comparison to Naive-Style LLM Usage}
\begin{table}
\caption{Ratio of correct patch placements and trust model broken (fix located within the sandbox) over 10 iterations.}
\footnotesize
\label{tab:function_role_accuracy}
\centering
\begin{tabular}{c|c|c}
\hline
Approach & \makecell{Placement OK} & \makecell{Trust model broken} \\
\hline
Naive GPT-4o-mini & 10\% & 0\% \\
Naive GPT5        & 20\% & 50\% \\
This work & 100\%  & 0\% \\
\hline
\end{tabular}
\end{table}

Table~\ref{tab:function_role_accuracy} presents the behavior of naive GPT-4o-mini and GPT5 prompting on this CIV. \texttt{Placement OK} is the fraction of runs in which the patch is placed at the policy-preferred location defined in Section~\ref{sec:priority} (i.e., \texttt{log\_table\_entry}). \texttt{Trust model broken} is the fraction of runs in which the model places the patch inside the sandboxed compartment, which violates the sandbox threat model because the untrusted sandbox cannot be assumed to sanitize its own outputs. Overall, naive prompting without compartment and trust-model context is unreliable. In contrast, our framework consistently selects the intended trusted-side patch site and never moves validation into the sandbox.

\begin{lstlisting}[float,
caption={GPT-5-generated patched function.},
label={lst:GPT5-generated-patch}]
static int log_table_entry(const apr_table_t *table,
    const char *name, char *buf, int buflen) {
  // ...
  const char *val;
  if (table == NULL || name == NULL) {
    return "-";
  }
  val = apr_table_get(table, name);
  return val ? val : "-";
  // ...
}
\end{lstlisting}

Listing~\ref{lst:GPT5-generated-patch} shows the patch produced by our baseline GPT5 prompting. To prevent the crash caused by dereferencing an invalid pointer, GPT5 adds a \texttt{NULL}-pointer check before passing the pointer to \texttt{apr\_table\_get}. However, the failure mode in this CIV is not limited to \texttt{NULL}: the pointer can be non-\texttt{NULL} yet unmapped. As a result, this patch is only a partial fix. In contrast, our repair in Listing~\ref{lst:apache-fix} introduces an additional check (via an oracle such as \texttt{is\_pointer\_mapped}) to reject unmapped pointers before dereference. This still does not address the harder case where the pointer is mapped but refers to a corrupted \texttt{apr\_table\_t} object. Handling mapped-but-corrupted structured state is left as future work. Possible directions include synthesizing stronger structural validation for the referenced data structure or escalating ambiguous cases for human review.

\section{Progress So Far and Evaluation Plans}

The framework is operational to a degree where we can obtain the results presented in Section~\ref{sec:case-study}.
Considering the architecture on Figure~\ref{fig:overview}, the patch generation component has been developed, including both the CIV and call stack analyses, which output feeds into the prompt generator.
The patch generation currently integrates with the GPT-4o-mini and GPT5 models, however that connection is made generic, and we plan to explore other models in a near future.
The ConfFuzz CIV fuzzer is integrated in the framework, and we are in the process of finalizing the validator component.

An important next step is the evaluation of our framework on a wide set of applications, trust models, and CIVs.
To that aim we plan to compare our work to a baseline, naive use of LLMs without a compartmentalization-aware prompt, as well as competitors in the form of existing APR frameworks~\cite{APR1, APR2, bouzenia2024repairagent}.
We plan to focus in particular on APR approaches targeting security vulnerabilities~\cite{APR_SEC_SOK, KULSUM, AIBMC}.
Regarding the patch generation step of our framework, beyond using LLMs directly, we also plan to explore leveraging AI agents~\cite{AGENTS} that build upon LLMs to interact automatically with a development environment and coordinate complex software engineering tasks with the goal of addressing CIVs.

The evaluation will be run over the large ConfFuzz~\cite{lefeuvre2022assessing} dataset of 600+ CIVs.
The ability of our framework and its baselines/competitors to address CIVs will be assessed both automatically by fuzzing and CIV reproduction, as well as manually by experts.

We anticipate the evaluation will highlight areas of improvement for our approach given its current state.
In particular, our current definition for the ideal CIV patch location has not been validated against a large amount of CIVs, and it may evolve in the future.
The classification of functions involved in a CIV's payload data flow that we propose in \ref{sec:stack-classification} should allow us to explore different strategies.
Another area of improvement we have not explored yet regards how to patch an application to handle the failure of interface security checks.

\section{Related work}

Compartmentalization~\cite{PRIVMAN, PRIVTRANS} has been the subject of many studies over the past two decades~\cite{lefeuvre2025sok}, however the problem of CIVs have seldom been explored.
The concept was first coined by ConfFuzz \cite{lefeuvre2022assessing}, an in-memory fuzzing framework \cite{coppik2019memfuzz, liu2024fuzzinmem} designed to discover vulnerabilities at compartment boundaries.
By mutating communication payloads that traverse trust boundaries \cite{luk2005pin}, ConfFuzz systematically stresses interface semantics and exposes CIVs.
Other works aim detect CIVs using static analysis~\cite{CIVSCOPE}.
Studies prior to ConfFuzz studied specific types of interface vulnerabilities that qualify as CIVs, including confused deputy~\cite{CONFUSED_DEP}, type confusion~\cite{TYPE_CONFUSION}, Iago~\cite{IAGO} or dereference under influence~\cite{DUI} attacks.

Automatic Program Repair (APR) \cite{zhang2023survey} has evolved substantially over the past decade, spanning rule based \cite{mechtaev2016angelix}, search based \cite{le2011genprog}, template based \cite{liu2019tbar}, semantic based \cite{nguyen2013semfix}, and learning based \cite{chen2019sequencer, zhang2023survey} approaches, and most recently transitioning toward LLM-based repair systems \cite{bouzenia2024repairagent} that leverage pretrained foundation models \cite{xia2023automated}. These models implicitly encode repair templates, bug patterns, and partial semantic reasoning capabilities, reshaping the repair landscape beyond traditional syntactic transformations.

Recent evaluations have demonstrated the efficacy of LLM-based APR. \cite{xia2023automated} shows that modern pretrained language models outperform most classical APR tools across multiple benchmarks. The study evaluates both decoder-only architectures \cite{radford2019language} (e.g., GPT-family models \cite{lajko2024automated}) and encoder-decoder architectures \cite{vaswani2017attention} under different repair formulations, including full function regeneration and localized infilling, with Codex \cite{fan2023automated} achieving the strongest empirical performance.
Codex \cite{fan2023automated} introduced the HumanEval benchmark to evaluate code synthesis and repair capabilities and proposed new evaluation metrics to more accurately capture functional equivalence beyond textual matching. In related work, \cite{lajko2024automated} evaluates one-line fixes generated by GPT-2, GPT-3, and Codex, comparing LLM-generated patches against developer authored patches. Results show that Codex, fine-tuned for code generation, consistently achieves the highest accuracy across top-k repair candidates.

Existing APR tools do not target CIVs and treat programs as monolithic pieces of trust.
This lack of compartmentalization knowledge is the gap we aim to bridge in this study.
\section{Conclusion}
Although software compartmentalization holds many promises, cross-compartment interfaces remain a major source of vulnerabilities that undermine its security benefits.
This early investigation observes that existing APR techniques and general-purpose/code-centric LLMs are inadequate as-is for repairing such issues.
To fill this gap and assess the suitability of LLM-based APR approaches at fixing compartment interface vulnerabilities, we introduce a framework coupling into a feedback loop a LLM-based patch generator made compartmentalization-aware with the help of two CIV classification techniques, and a compartment interface fuzzer producing CIVs and validating patches.
Initial results on a representative vulnerability highlight the promise of this approach over naive LLM use, and point to several avenues for advancing automated hardening of compartment interfaces.

\section*{Acknowledgments}

We thank the anonymous reviewers for their comments and insights.
This work is supported in part by Innovate UK grant 10164504 (MicrOS), and the UK Engineering and Physical Sciences Research Council grants EP/V012134/1 (UniFaaS), EP/V000225/1 (SCorCH), and EP/X015610/1 (FlexCap).

\bibliographystyle{ACM-Reference-Format}
\bibliography{bib}

\end{document}